\documentclass[a4paper,11pt]{article}

\usepackage{jinstpub} 
\usepackage{tabularray}
\usepackage{tabularx}
\usepackage{hhline}
\usepackage{lineno}
\usepackage{tikz}
\usepackage{siunitx}
\usepackage{svg}

\title{\boldmath The SNEWS~2.0 Alert Software for the Coincident Detection of Neutrinos from Core-Collapse Supernovae}

\author[a]{M. Kara,}
\author[b]{S. Torres-Lara}
\author[c]{A. Baxter-Depoian,}
\author[d]{S. BenZvi,}
\author[e]{M. Colomer Molla,}
\author[f]{A. Habig,}
\author[g]{J.P. Kneller,}
\author[h]{M. Lai,}
\author[c]{R.F. Lang,}
\author[c]{M. Linvill,}
\author[c]{D. Milisavljevic,}
\author[i]{J. Migenda,}
\author[c]{C. Orr,}
\author[j]{K. Scholberg,}
\author[k]{J. Smolsky,}
\author[l]{J. Tseng,}
\author[m]{C.D. Tunnell,}
\author[n]{J. Vasel,}
\author[o,p]{A. Sheshukov}

\affiliation[a]{Institute for Astroparticle Physics, Karlsruhe Institute of Technology, 76021 Karlsruhe, Germany}
\affiliation[b]{Department of Physics, University of Houston, Science and Research Building 1, 3507 Cullen Blvd Room 617, Houston, TX 77204, USA}
\affiliation[c]{Department of Physics and Astronomy, Purdue University, 525 Northwestern Ave., West Lafayette, IN 47907, USA}
\affiliation[d]{Department of Physics and Astronomy, University of Rochester, 500 Wilson Blvd., Rochester, NY 14627, USA}
\affiliation[e]{Université Libre de Bruxelles, 1050 Bruxelles, Belgium}
\affiliation[f]{Department of Physics and Astronomy, University of Minnesota Duluth, 1023 University Dr., Duluth, MN 558012}
\affiliation[g]{Department of Physics, NC State University, Raleigh, NC 27695}
\affiliation[h]{Department of Physics and Astronomy, University of California Riverside, Riverside, CA 92521, USA}
\affiliation[i]{e-Research, King's College London, London, UK}
\affiliation[j]{Department of Physics, Duke University, Durham, NC 27708, USA}
\affiliation[k]{Department of Physics, Massachusetts Institute of Technology, 77 Massachusetts Ave, Cambridge, MA 02139, USA}
\affiliation[l]{Department of Physics, Oxford University, Oxford OX1 3RH United Kingdom}
\affiliation[m]{Department of Physics and Astronomy, Rice University, Houston, TX 77005, USA}
\affiliation[n]{520 S Walnut St \#1444, Bloomington, IN 47401}
\affiliation[o]{Joint Institute for Nuclear Research, Dubna, Moscow region 141980, Russia}
\affiliation[p]{Institute for Nuclear Research of the Russian Academy of Sciences, Moscow, 117312, Russian Federation}

\emailAdd{melih.kara@kit.edu}
\emailAdd{sptorres@cougarnet.uh.edu}

\abstract{The neutrino signal from the next galactic core-collapse supernova will provide an invaluable early warning of the explosion. By combining the burst trigger from several neutrino detectors, the location of the explosion can be triangulated minutes to hours before the optical emission becomes visible, while also reducing the rate of false-positive triggers. To enable multi-messenger follow-up of nearby supernovae, the SuperNova Early Warning System~2.0 (SNEWS~2.0) will produce a combined alert using a global network of neutrino detectors. This paper describes the trigger publishing and alert formation framework of the SNEWS~2.0 network. The framework is built on the HOPSKOTCH publish-subscribe system to easily incorporate new detectors into the network, and it implements a coincidence system to form alerts and estimate a false-positive rate for the combined triggers. The paper outlines the structure of the SNEWS~2.0 software and the initial testing of coincident signals.}


\begin{document}
\maketitle
\flushbottom

\section{SNEWS Introduction}

The Supernova Early Warning System (SNEWS) is a global network of detectors that aims to provide early warning of nearby core-collapse supernova (CCSN) explosions.  The neutrino signal from a core-collapse supernova, which significantly increases at the point of explosion, provides a valuable early warning for such celestial events \cite{Scholberg:2012id}.\footnote{Neutrinos are emitted in vast quantities from a star as it collapses, reaching a peak coinciding with the supernova explosion. This signal precedes the optical visibility of the supernova, making it crucial for early detection.} SNEWS aims to quickly distribute the early warnings across a network of neutrino and dark matter detectors. Each detector is independently sensitive to CCSN neutrinos. Taken together, the combined signal provides richer information than any single experiment could achieve alone.

In its original form, SNEWS \cite{Antonioli:2004zb} started in 2001 and has been fully operational since 2005, searching for a coincident signal between the participating experiments. However, its current implementation demands a low false alarm rate from its participants and relies on outdated hard-to-maintain code. To address these limitations, a new version of SNEWS, called SNEWS~2.0, was developed \cite{SNEWS:2020tbu}. This new system not only detects time coincidences between signals sent by different detectors but also provides a new functionality to combine the signals, such as significance stacking and directionality.   

This paper focuses on two Python packages: the SNEWS Publishing Tools (\texttt{snews\_pt}), a client-side library designed to communicate observation messages, and the SNEWS Coincidence System (\texttt{snews\_cs}), a server-side program responsible for monitoring and identifying coincidences among the received messages. Developed to facilitate user participation, these packages play crucial roles in managing data interaction and enhancing the efficiency of the SNEWS network.

\subsection{SNEWS 1.0}

The original SNEWS network \cite{Antonioli:2004zb} started at a workshop in 1998, where neutrino experimentalists, supernova theorists, and astronomers sought to exploit the capability of existing detectors to detect the early galactic CCSN signature provided by neutrinos.  
Neutrino experiments are sensitive only to nearby (in the Milky Way) core-collapse SNe, estimated to occur 1.63$\pm$0.46 times per century \cite{Rozwadowska:2021lll}. Given the rarity of such events, individual experiments exercise caution in declaring the detection of signals, mindful of the significant implications of a potential discovery error.

Human validation at a single experiment might take hours, potentially eliminating the valuable warning provided by neutrinos. However, the occurrence of signals across multiple experiments concurrently significantly diminishes the likelihood of false alarms. This allows an automated coincidence system to issue a prompt alert to the world, maximizing the time dedicated to observing the electromagnetic fireworks.

This first instance of SNEWS was written in the C programming language using raw network sockets, with an API and command line client provided so that participating experiments could send a very simple TCP datagram over unix network sockets to a central server (running at Brookhaven National Lab) as well as a backup (at INFN Bologna).  The datagram contained the time of the observation and several optional fields that in practice never got used.  The server kept track of the experimental alerts and, if two or more arrived within \SI{10}{\second}, would report that to an email list of interested parties.  Currently, there are nearly 7,000 subscribers to the mailing list, which, given the fact that a nearby CCSN has not occurred since 1987, might be the lowest-traffic mailing list ever to be on the Internet. 
SNEWS came online in a test mode in 2001, and has been fully operational since July 1, 2005, making it one of the first and longest-running Multi-Messenger Astronomy (MMA) projects.  No CCSN has occurred in our galaxy since that time, nor has SNEWS reported a neutrino coincidence that would suggest one had.    Experiments restrict themselves to less than approximately one false alarm per week, which results in a Poisson probability of accidental coincidence of less than the actual SN rate (one per century).  

This network continues to operate and will continue to operate until the new SNEWS~2.0 is well established.  However, there are several reasons to upgrade.  First, the original code, written 25 years ago for now obsolete VMS, IRIX, and AIX computers, is hard to port to new experiments.  In addition to maintainability, the current proliferation of MMA projects has changed the community's perception of alert publishing.  When SNEWS started, one false alarm would have been the death of the project.  Now, people realize that to maximize results, false alarms can be tolerated, provided that confidence levels are provided so consumers can choose their own appropriate action.  Furthermore, with the increased size of modern experiments, the once statistically intractable problem of using neutrino arrival times to point back to an area in the sky is now possible, with a similar uncertainty to the LIGO/VIRGO gravitational wave skymaps \cite{Brdar:2018zds,Linzer:2019swe,Coleiro:2020vyj} that enabled the identification of electromagnetic counterparts. Finally, a wide range of automated follow-up observatories are now active and chasing after gamma-ray bursts, gravitational waves, and high-energy neutrinos.  Thus, the SNEWS~2.0 effort \cite{SNEWS:2020tbu} to address these changes.

\subsection{SCiMMA and hop-client} 

To establish communication with detectors and members of the astronomy community, SNEWS uses Scalable Cyberinfrastructure to support Multi-Messenger Astrophysics (SCIMMA) \cite{scimma-hopskotch} architecture, which provides a scalable data stream backend called HOPSKOTCH along with a Python API \textit{hop-client}. 

 HOPSKOTCH is a cloud-based instance of a Kafka \cite{Kafka} data stream designed to support MMA initiatives by providing users with low-latency communication through publish and subscribe protocols.
 Kafka data streams or channels act as real-time data pipelines, allowing users to distribute their data across a secure network. 
 For instance, a user can publish a message through a Kafka channel using the \textit{hop-client} API, enabling a Python implementation of Kafka channels for publishing (sending) and subscribing (receiving) data. Users subscribed to that channel will receive the messages posted there. The hop-client has been an integral tool for the SNEWS 2.0 upgrade \cite{SCiMMA:2021kab}, thanks to its intuitive functionality and the continuous support from the SCIMMA developers. 

The ensuing Sections 2 and 3 provide a detailed overview of the SNEWS Publishing Tools (\texttt{snews\_pt}) and the SNEWS Coincidence System (\texttt{snews\_cs}), which are critical software components designed to enhance the robustness and responsiveness of the SNEWS network. By optimizing the interactions between detectors and the central server, these tools facilitate rapid and reliable communication across the global array of neutrino detectors, ensuring that potential supernova alerts are processed and disseminated with high efficiency and accuracy. This discussion highlights the operational mechanics and functionalities of \texttt{snews\_pt} and \texttt{snews\_cs}, underscoring their pivotal contributions to the multi-messenger astronomy objectives of SNEWS 2.0.

\section{SNEWS Publishing Tools}

The SNEWS Publishing Tools, \texttt{snews\_pt}, is the Python front-end software designed to enhance the interaction between detectors and the SNEWS servers. The tools enable detectors to easily publish observation messages and their data-taking status (referred to as Heartbeats), as well as subscribe to alerts issued by the server. These alerts are triggered by the SNEWS Coincidence System, \texttt{snews\_cs}, software running on the server described in the next section.

The \texttt{snews\_pt} provides flexibility by offering its users a choice of using either a Python API or a command-line interface. The software encompasses three main functionalities: subscribe to alerts, publish heartbeats, and transmit observation messages. In addition to these core functionalities, the software offers utility functions called ``commands'', which fall into two distinct categories. The first category consists of commands designed primarily for developers to facilitate testing. An example of such a command is a \texttt{cache\_reset} request to clean its cache and initiate a fresh start, enhancing the testing experience. The second category includes commands that provide valuable information to users such as \texttt{test\_connection} to validate connection with the server. Further details on each functionality and the primary purposes of the existing commands are discussed in \ref{subsec:remote commands}.

The software also includes an ``environment'' file containing high-level configuration information, such as precise URL addresses for observation and alert channels deployed for various purposes. Consequently, users need only specify their intended use of the tools, as the functions can easily retrieve the necessary information at runtime. Including such a configuration file also enhances the flexibility for future production scenarios.

\subsection{Credentials}

Through the HOPSKOTCH system, users can obtain authorized access to specific channels within the SNEWS network, including read and write permissions. The SNEWS network relies on two distinct Kafka channels for exchanging messages. The first channel grants write access, allowing specifically authorized users affiliated with recognized member experiments to securely submit their observation messages and heartbeats. The second channel grants read access to users, serving as the conduit for the transmission of alert messages that prompt users to take necessary actions. This design enables users to stream their observations into a unified channel while simultaneously subscribing to incoming alert notifications from the other channel.

The SNEWS Coincidence System is the only participant allowed to read from the first channel and write to the second. This permission configuration ensures the efficient flow of critical information within the network.

\subsection{Publishing}
\label{subsec:publishing}

To facilitate the publication of experimental information, active SNEWS members can interact with the \texttt{messages} module, using its \texttt{SNEWSMessageBuilder} class. This class provides a flexible method for constructing and validating messages, enabling users to create messages tailored to various purposes. Different data tiers are discussed in \cite{SNEWS:2020tbu} and are also shown in Table \ref{tab:message content}. \texttt{snews\_pt} identifies the intended tiers based on the provided inputs, ensuring that messages are generated in suitable formats to accommodate different requirements. Furthermore, the system performs a comprehensive validation of the message input during construction, promptly identifying incorrectly formatted or missing inputs and notifying users. This not only secures accurate data processing but also helps prevent server crashes or stalls, thereby ensuring long-term stability and reliability of the system.

\begin{table}
\centering
\begin{tabularx}{\linewidth}{
  >{\centering\arraybackslash}m{2cm}
  |>{\raggedright\arraybackslash}m{1.5cm} 
  |>{\centering\arraybackslash}X 
  |>{\centering\arraybackslash}X 
  |>{\centering\arraybackslash}X
  |>{\centering\arraybackslash}X
  |>{\centering\arraybackslash}X 
}
                       & Format                  & Coincidence        & Significance       & Timing             & Heartbeats         & Retraction         \\
\hline
detector name          & str                     & \textbf{Required}  & \textbf{Required}  & \textbf{Required}  & \textbf{Required}  & \textbf{Required} \\ \cline{1-2}
initial neutrino time  & str                     & \textbf{Required}  & -                  & \textbf{Required}  & -                  & -                  \\ \cline{1-2}
machine time           & str                     & Optional           & Optional           & Optional           & Optional           & Optional           \\ \cline{1-2}
observation p-value    & float [0,1]             & Optional           & Optional           & Optional           & -                  & -                  \\ \cline{1-2}
p-values for time bins & list [float]            & -                  & \textbf{Required}  & -                  & -                  & -                  \\ \cline{1-2}
width of time bins     & float                   & -                  & \textbf{Required}  & -                  & -                  & -                  \\ \cline{1-2}
neutrino time series (histograms)  & list \newline[str (int)] & -             & -                  & \textbf{Required}  & -                  & -                  \\ \cline{1-2}
detector status        & str (ON|OFF)            & -                  & -                  & -                  & \textbf{Required}  & -                  \\ \cline{1-2}
retract latest         & int                     & -                  & -                  & -                  & -                  & \textbf{Required}  \\
\end{tabularx}
\caption{Detailed specifications of message arguments for the SNEWS network. This table categorizes each argument by required or optional status across different message tiers, highlighting how detectors should format their observational data for submission.}
\label{tab:message content}
\end{table}

The \texttt{SNEWSMessageBuilder} offers a comprehensive set of recognized arguments, shown in Table \ref{tab:message content}. The table provides an overview of required and optional arguments for different tiers within the SNEWS system\footnote{Due to the evolving nature of the message details, especially for the significance and timing tiers, the specifics provided in Table 1 are subject to change. For the most current and detailed descriptions of the message parameters, readers are encouraged to consult the online documentation. The online documentation can be accessed at \url{https://snews-publishing-tools.readthedocs.io/en/latest/}.}. The first two columns depict the argument and its anticipated format, while the subsequent columns categorize these parameters according to distinct message tiers. The rows detail the fields available to the user in each message tier. 
Users can also specify if the message is meant for testing by passing an \texttt{is\_test=True} argument. In these cases software skips several validation steps, such as expecting the neutrino times to be within the last 48 hours, that would typically apply in real-case scenarios. This feature enables users to test their messages at random times, enhancing message validation and testing capabilities.

Users can also provide the \texttt{machine\_time} information, which indicates the exact International Organization for Standardization (ISO) time when the data were read by the experiment's machine. Additionally, any times provided, such as neutrino time, machine time, and neutrino time series (if given as strings), need to be in ISO format. If the neutrino time series is not a list of strings, the software expects a list with nanosecond precision integers from the initial neutrino time. The software also automatically appends the \texttt{sent\_time} information upon execution. This capability allows the server to interpret latencies between the read and send times of the experiments, providing valuable insights into the detector's status and data flow.

Furthermore, the tool exhibits versatility by accepting any additional key-value pairs introduced by users as metadata. These metadata are directly recorded into the messages. This functionality provides adaptability and facilitates the transfer of unforeseen yet valuable data.

\subsection{Subscribing}
\label{subsec:pt_sub}
\texttt{snews\_pt} offers a user-friendly and efficient method for subscribing to the SNEWS alert channel. Notably, this subscription process is accessible to a wide user base, extending beyond member experiments to include any individuals or observatories with Kafka credentials. These users can subscribe and actively monitor the incoming alerts using the \texttt{Subscriber} class, accessible through the API or the terminal.

The alerts received are not only presented to users for immediate review but are also automatically stored in a designated local output folder in JSON file format. This storage facilitates subsequent in-depth investigation and analysis of the alerts received.

For users who seek to respond promptly to issued supernova alerts and take actions based on the alert information, \texttt{snews\_pt} integrates smoothly with follow-up scripts. Upon receiving an alert, the subscriber quickly generates a JSON file containing the alert's content. Subsequently, it automatically triggers the execution of a pre-defined follow-up script, passing the file's absolute path as an input. Users can add just two lines to their custom script to read the JSON file's content as a dictionary, allowing them to use the information for follow-up tasks. This automated execution of the custom script creates a fully integrated workflow, enabling users to respond efficiently to supernova alerts and perform subsequent analyses without the need for manual intervention.

\subsection{Remote Commands}
\label{subsec:remote commands}
While the primary functionalities of the \texttt{snews\_pt} revolve around subscribing to alerts and broadcasting observations and heartbeats, the tools also offer a set of remote commands that can prove valuable in specific scenarios. These remote commands are formatted as message blobs with key-value pairs and are recognized as ``requests'' by the server upon reception. Some of these commands are particularly designed for development purposes, aiding in easier testing.

Remote commands that are currently available are listed in Table \ref{tab:remote commands}. The \texttt{reset\_cache} command, allows users to instruct the server to clean its cache memory. This action initiates a new test environment, preventing potential coincidences with previously submitted messages and ensuring accurate results during testing.

\begin{table}[!h]
\begin{tabular}{l|l|l}
                & Used by   & Purpose\\
\hline
\texttt{reset\_cache}     & Developers & In the testing environment, drop the cache and start fresh. \\
\texttt{test\_connection} & All users & Confirm that your messages are read by the server.                \\
\texttt{get\_feedback} & All users & Request Email Feedback on Heartbeats from last 24h.             \\
\end{tabular}
\caption{Summary of the currently available remote commands.}
\label{tab:remote commands}
\end{table}

The \texttt{test\_connection} command lets users validate their connection to the server. Upon execution, the function generates a message containing the execution timestamp, detector name, and a status indicating that the message is being sent (\texttt{"status": "sending"}), sending it to the \textit{observation channel}. Subsequently, the function subscribes to a distinct \textit{connection test channel}, anticipating a confirmation message from the server. Upon reception, the server generates a new message, retaining all details and updating the status to indicate that the message has been received (\texttt{"status": "received"}). This modified message is then published to the \textit{connection test channel}. \texttt{snews\_pt} then compares and validates the received message, establishing a loop between the \textit{observation channel} (with user write access) and the \textit{connection test channel} (with user read access). This iterative process not only completes the communication loop between the user and the server but also serves to validate the operational status of the server, confirming its active and functional state.

The \texttt{get\_feedback} command lets users request feedback on heartbeats submitted within the last 48 hours. In the server's ongoing process, heartbeats are retained for a brief period of 48 hours before automatic removal. The function prompts users to input email address(es) to receive the feedback. Upon receiving the request, the server checks that the provided email address(es) are registered as contacts for the associated experiment. When the request originates from a recognized address, the server generates a feedback image, containing essential information such as the latencies and interpreted frequencies, and sends an email to the recognized email addresses. This feedback not only serves as a tool for users to assess the status of their systems but also provides insights into latencies, contributing to an understanding of the overall system dynamics. 

\begin{figure}[!h]
    \centering    \includegraphics[width=\textwidth,height=\textheight,keepaspectratio]{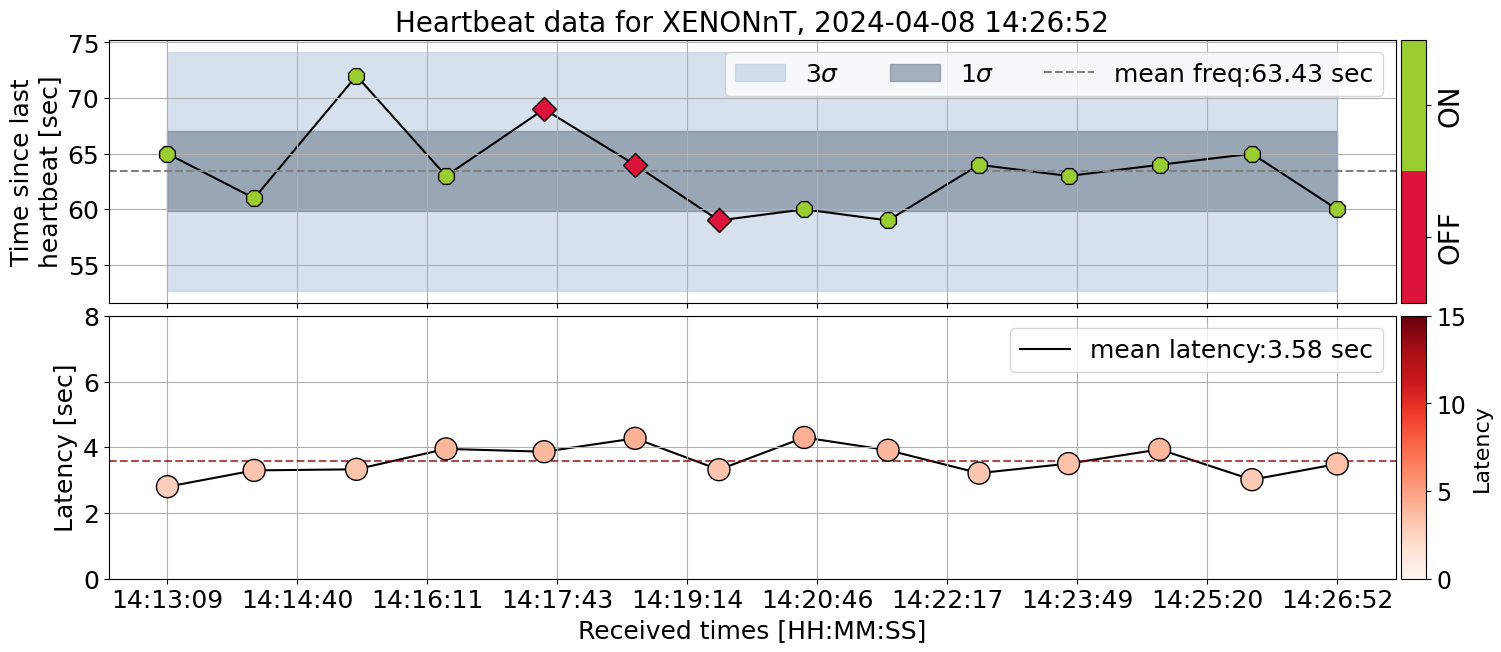}
    \caption{A simulated example image illustrating the feedback from the server containing information of several heartbeats that were registered within about 15 minutes.}
    \label{fig:feedback}
\end{figure}

Figure \ref{fig:feedback} presents an example feedback image, simulated to illustrate heartbeats registered within approximately 15 minutes from XENONnT experiment \cite{XENON:2024wpa}. The top panel visually represents the frequency of the registered heartbeats, with online detector status beats plotted in green and offline status in red. The mean frequency, along with the one- and three-standard deviation bands, is highlighted for reference. Additionally, the second panel shows the latency, defined as the time between the user submitting a heartbeat (\texttt{sent\_time}) and the server receiving that message (\texttt{received\_time}), providing insights into the time delays experienced by each experiment for further analysis and optimization.

These remote commands extend the functionality of the \texttt{snews\_pt}, providing users with advanced testing options and ensuring smoother communication with the server during various development tasks.

\subsection{Retraction and Skipped Heartbeat Warnings}

\texttt{snews\_pt} provides users with the capability to retract previously submitted messages, offering a mechanism to rectify errors or refine analyses and mitigate the risk of generating false alarms. This process mirrors the submission of a new observation message using the \texttt{MessageBuilder} detailed in Section \ref{subsec:publishing} and presented in Table \ref{tab:message content}. To initiate a retraction, the user simply specifies how many of their most recent messages they intend to retract. The software interprets this as a retraction request. Subsequently, the server identifies and removes the latest $N$ messages submitted by the requesting experiment from its cached data. Following this removal, the coincidence logic is re-initialized. If an alert was previously triggered, an updated message is broadcasted, reflecting the absence of the now-retracted detector in the system. The system triggers an updated alert even if only a single detector remains after the retraction of the second detector that originally contributed to the coincidence and triggered the initial alert.

Heartbeats play a crucial role in the overall functionality of the early warning system. Through the heartbeats, the server running the SNEWS Coincidence System can actively monitor and alert users in case of a lost connection at any given point. While the server anticipates receiving heartbeats from all detectors, it does not impose a specific frequency requirement. Once the server accumulates enough heartbeats, it calculates the heartbeat frequency and assigns a value of $\mu \pm \sigma$ to the corresponding experiment. With the continuous accumulation of heartbeats, statistical accuracy improves, and the expected frequency is dynamically updated. With the established frequency, the server anticipates receiving the next heartbeat within this time interval. The server conducts this verification process every minute, and if no new heartbeat is registered after the $\mu+3\sigma$ time has elapsed, it generates a warning email. This email includes the collected statistics and is sent to the pre-registered responsible email addresses associated with that specific experiment. Upon the absence of new heartbeats and the generation of a warning email, the checks for this particular experiment are temporarily halted. The system resumes the frequency computation once new heartbeats start accumulating, ensuring that the calculations are based on the most recent data. This adaptive approach allows the system to adjust to changing conditions dynamically and provides accurate frequency assessments for effective monitoring and alert handling.

\section{SNEWS Coincidence System}
The SNEWS Coincidence System, \texttt{snews\_cs}, serves as the central communication hub of SNEWS, processing Coincidence Tier and Heartbeat messages and generating prompt alerts when specific alert conditions are met. 
This system was developed using Python, leveraging the Pandas library for data manipulation and online data storage, SQLite for long-term data storage, and a hop-client API for seamless integration with the SNEWS SCIMMA broker. This enables the system to receive messages from participating detectors and distribute alerts to subscribed users through the SNEWS alert broker.
\texttt{snews\_cs} contains of two classes, the \textit{CacheManager} and the \textit{CoincidenceDistributor}.

The \textit{CacheManager} class establishes the coincidence cache and organizes data within it. 
The primary objective of this class is to store incoming messages in an online cache and determine coincident instances. Furthermore, this class was designed to effectively handle staggered signals by implementing coincident \textit{subgroups}, or subsets of the messages in the cache.  To achieve this, the \textit{CacheManager} generates subgroups when a new message falls outside the coincidence time range of existing subgroups. This approach allows for the formation of coincidences between distinct signal sets, facilitating the identification and potential retraction of any erroneously submitted signals. Incorporating coincident subgroups enhances the system's ability to discern genuine coincidences from chance occurrences, improving accuracy and reliability. Additionally, the \textit{CacheManager} class incorporates robust checks to prevent the formation of redundant subgroups, ensuring that the cache is free of superfluous data. 

The \textit{CoincidenceDistributor} class creates a data stream to the SNEWS detectors through the hop subscription method.
This data stream establishes real-time communication between SNEWS and participating detectors, where Coincidence Tier messages can be sent to the Coincidence System.
During operation, all incoming messages are passed through an authentication process, where any malformed messages or messages with missing data are discarded, preventing instabilities during the run time.
Following authentication, the contents of each message are validated, confirming that they satisfy the criteria for inclusion within the coincidence cache. 
Once validated, messages are passed to the \textit{CacheManager}, populating the active coincidence cache.
Whenever a new message is processed, the \textit{CacheManager} is prompted to check for coincidences. 
If a coincidence is detected, then an alert message is created and sent across the network, informing all participating members that a coincident event was observed.

\subsection{Coincidence Data Flow}
The Coincidence System processes messages sequentially. Once received, a message is passed through a quality-checking method, which acts as a redundant filter for the \texttt{snews\_pt}'s validation system. 
This step is crucial if any detector's messages are inadvertently transmitted to the server.
If a message meets the standard, it is stored in the cache as a Pandas data-frame object. The cache retains all coincident messages for 48 hours.
If the cache is empty, the incoming message initializes the first sub-group and establishes the initial neutrino time \(\left( t_{\nu,m} \right)\).
A detector's neutrino time corresponds to the first neutrino signal it observes during a supernova flux event.
When a subsequent message arrives, its \(t_{\nu,m}\) is compared to the current initial neutrino time of a sub-group \(\left( t_{\nu, sg} \right)\). 
If the incoming message has a neutrino time difference \( > \SI{10}{sec}\), then it is stored in the cache as an initial message for a new sub-group. 
Conversely, if the neutrino time difference between \(t_{\nu,m}\) and \(t_{\nu, sg}\) is \(\leq \SI{10}{sec}\), the message is assigned to the sub-group to which it coincides.
 Finally, if a message arrives with an earlier neutrino time than \(t_{\nu, sg}\) and still falls within the coincidence window of the sub-group,  it is designated as the new initial message.
 
 Regarding the alert distribution, once a sub-group accumulates two or more coincident messages, an alert is dispatched through a public Kafka output channel.
 As mentioned in Sec. \ref{subsec:pt_sub} \texttt{snews\_pt} provides subscription and publishing methods, enabling the alerts to be published to different Kafka channels and in different formats.
 These alerts can be sent to participating detectors, ground-based telescopes, gravitational wave observatories, and amateur astronomers. 
 Furthermore, the early warning alerts are distributed through the original SNEWS mailing list and as a GCN alert \cite{gcnalerts} alert (which is then picked up by the SNEWS mobile app). 
 A more detailed alert is sent to participating detectors, informing them of the time window of a supernova search. 

\begin{figure}
    \centering
    \includegraphics[width=\textwidth,height=\textheight,keepaspectratio]{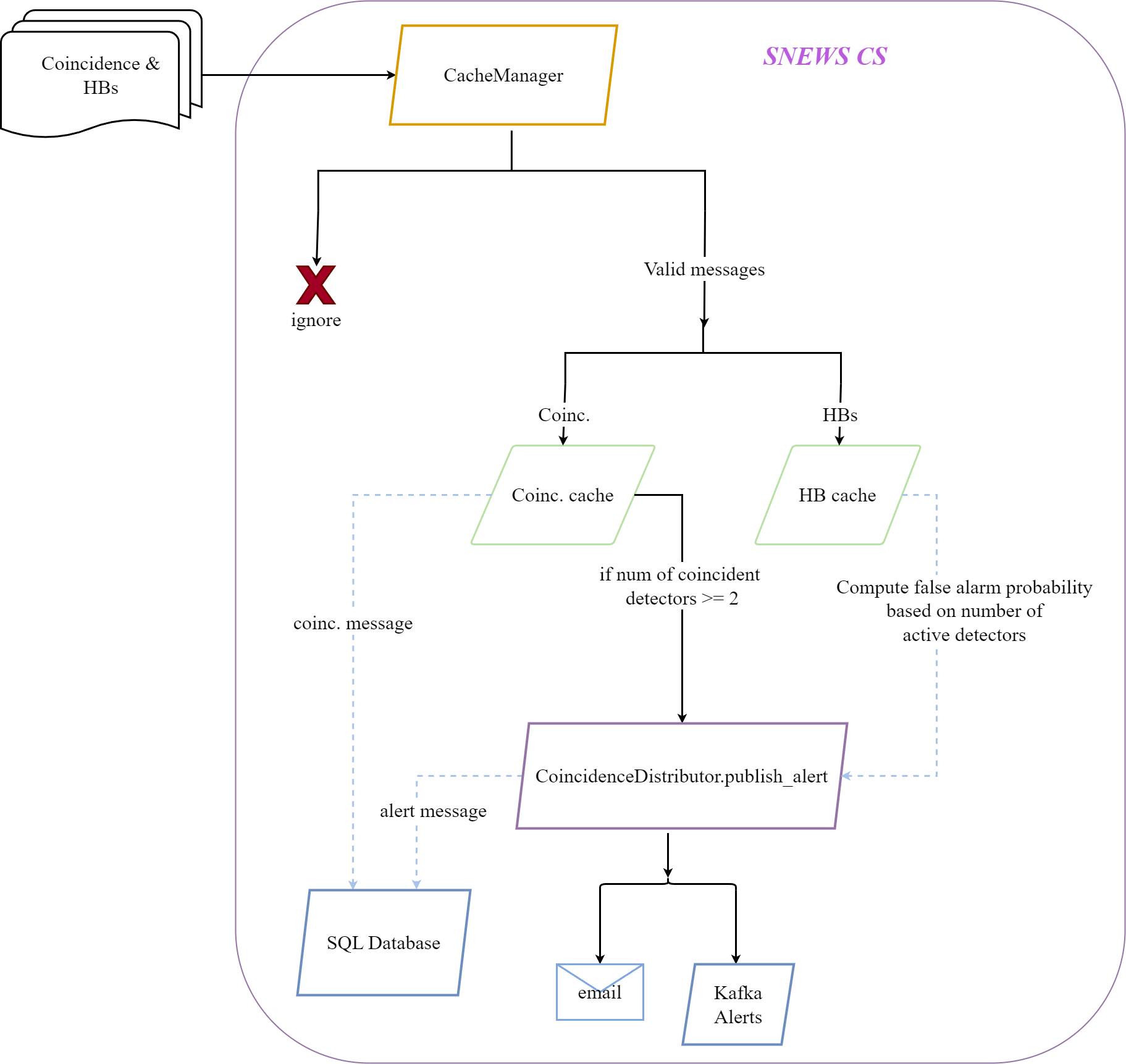}
    \caption{A diagram showing the data flow in the Coincidence System concerning the coincidence tier messages and heartbeats.}
    \label{fig:dataflow}
\end{figure}

 Assuming a sub-group continues to receive more coincident messages, a subsequent alert will be sent whenever a new message is received, updating the alert metrics as new information is received. An overview of the SNEWS CS data flow is shown in Fig. \ref{fig:dataflow}. 

\subsection{SNEWS Operations}

The SNEWS Coincidence System server exists as an Ubuntu-based Apptainer \cite{Apptainer}
container running on a Linux host. 

The benefits of containerization are well known; 
portability, isolation, agility, and ease of management, to name a few. However, the primary driver in this choice for SNEWS is the ability to run a global network of 
distributed, redundant servers simply from a single code base. On the host, \texttt{systemd} unit files supervise the running processes, restarting them if needed, with custom supervisory code providing status messaging via Slack. In addition to the supervisory code, the Zabbix monitoring system is used internally to provide notifications and hardware performance metrics for the host computer to systems administrators.

Separate production and development containers are
run in parallel.  The production and development containers typically run the same code base unless the development container is testing new code.  The primary difference between production and development is which SCiMMA HOPSKOTCH channel that the instance is subscribed to.  In the case of the development container, the hop client would be subscribed to dedicated testing channels to ensure readiness and availability.  The production container is left to be production worthy and should only alert when necessary.

Currently, a Grafana dashboard is under development to provide useful information at a glance to researchers and system administrators alike. This dashboard will display data from recently-detected coincidences and provide comprehensive system latency visualizations from individual detectors to \texttt{snews\_cs}.  We intend to provide a web-page that will display coincidences for the general public to view.

\section{Firedrills}

SNEWS conducts periodic distributed tests called ``firedrills'' to test that both \texttt{snews\_pt} and \texttt{snews\_cs} are operating as intended, to validate the pointing capabilities, and to test system updates. 

Since the galactic supernova event rate is extremely low it is of the utmost importance that SNEWS~2.0 is capable of working under load, making sure that the code is bug-free, host servers are working as intended, and that the Hop broker allows sending the corresponding data. 

For the development team, firedrills allow a wider audience to give feedback on the \texttt{snews\_pt} and the alert protocols of the \texttt{snews\_cs}. 
Furthermore, edge cases, such as staggered coincidences or updating an alert when the detector retracts a message, are tested during firedrills.
Frequent testing helps participating SNEWS members familiarize themselves with the deployment of \texttt{snews\_pt}.

 A typical firedrill involves SNEWS members participating in a virtual call to test various alert scenarios. These scenarios include a simple coincidence, where all detectors register the supernova flux simultaneously, a staggered coincidence involving a subset of detectors with delayed data transmission, and a coincidence with retractions and detector updates.
 
 Firedrills provide a crucial learning experience for the development team to enhance the software and refine the implementation of these systems within SNEWS.
 Notably, the most significant updates to both \texttt{snews\_pt} and \texttt{snews\_cs} were implemented after thorough feedback by the SNEWS community following a firedrill. For instance, during a firedrill in October 2022, participating SNEWS members contributed the following improvements: changed the \texttt{snews\_pt} logic to exclude testing messages for detectors; established a proper naming convention for detectors; enabled messages to include time uncertainty of the neutrino signal; and updated the alert statistics following an update or retraction.

\section{Conclusion} 

By leveraging a more flexible and versatile code base, SNEWS member experiments can more easily publish and receive alerts to potential galactic supernovae. SNEWS~2.0 has developed a Python-based framework based on SCIMMA’s HOPSKOTCH system for communications with neutrino detectors and the world. This Python framework consists of two parts: \texttt{snews\_pt} allows experiments to communicate their potential supernova alerts, and \texttt{snews\_cs} receives those alerts, forming coincidence alerts if two or more unique signals are within a $10$-second window. Once coincident signals are detected, a prompt alert is published to the MMA community and the public. This framework implements a wide range of plugins to receive the alert and distribute it in various required formats, including email, GCN alerts, or directly integrating with \texttt{snews\_pt} to feed custom code back into an experiment or observatory. Each software is designed to be sustainable and scalable, allowing SNEWS~2.0 to incorporate additional functionalities such as pointing or source significance stacking.  With SNEWS~2.0, the MMA community will receive an alert prompt enough to adapt the orientation of telescopes and antennas accordingly, assuring the most sensitive detection of the burst, providing a unique chance to learn about currently un-observed details of the explosion process. Additionally, the capability to reduce the uncertainty on the start of the burst can significantly aid the chances to detect gravitational waves from a CCSN \cite{Nakamura:2016kkl}.  

\section*{Acknowledgement}

This work is supported by the National Science Foundation ``Windows on the Universe: The Era of Multi-Messenger Astrophysics'' program: ``WoU-MMA: Collaborative Research: Advancing the SuperNova Early Warning System" through grants 2209449, 2209451, and 2209534 as well as NSF AST-2206532. We also acknowledge the support from the Science and Technology Facilities Council (STFC), United Kingdom.

\bibliography{bibliography}

\end{document}